\documentclass[conference]{IEEEtran}
\pdfoutput=1
\ifCLASSINFOpdf

\else

\fi

\usepackage[utf8]{inputenc} 
\usepackage[T1]{fontenc}    
\usepackage{hyperref}       
\usepackage{url}            
\usepackage{booktabs}       
\usepackage{amsfonts}       
\usepackage{nicefrac}       
\usepackage{microtype}      

\usepackage{graphicx}
\usepackage{amsmath}
\usepackage{algcompatible}
\usepackage{algorithm}

\begin{document}
%
\title{Bias reduction of peer influence effects with latent coordinates and community membership}

\author{\IEEEauthorblockN{Daniel Rajchwald}
\IEEEauthorblockA{ Harvard University\\
 Cambridge, MA 02138\\
rajchwa1@gmail.coml}
\and
\IEEEauthorblockN{Natasha Markuzon}
\IEEEauthorblockA{Draper Laboratory\\
Cambridge, MA 02139\\
nmarkuzon@draper.com}}


\maketitle

\begin{abstract}
The importance of peer influence on consumer actions plays a vital role in marketing efforts. However, peer influence effects are often confounded with latent homophily, which are unobserved commonalities that drive friendship. Understanding causality has become one of the pressing issues of current research. We present an approach to explicitly account for various causal influences. We implement a simulation framework to show the effectiveness of two latent homophily proxies, latent coordinates and community membership, in improving peer influence effect estimates on game downloads in a Japanese social network website. We demonstrate that latent homophily proxies have no significant improvement in peer influence effect bias in the available website data. 
\end{abstract}

\section{Peer influence and latent homophily}
Advertising companies have long been interested in the role of peer influence in social networks. Depending on how much consumers base their decisions on peer activity, advertisers can adjust their strategies accordingly [4][9]. Inferring the effect of peer influence is difficult because peer influence is confounded with user homophily. In other words, consumers may be influenced by their peers' actions or may be inherently prone to make the same decisions as their peers since people often befriend others similar to themselves.

Controlling for confounding factors and better identifying real causalities is an active area of research. Van den Bulte and Lilien [10] show how peer influence contributes to new drug adoption among doctors in the same social network. However, after controlling for drug company marketing, they find no significant peer influence effects on drug adoption. Another approach to identifying peer influence effects is to include observed covariates in regressions or matching. Observed features could also be correlated with homophily so using them can reduce bias of peer influence effects due to homophily. Aral et al. [2] use propensity score matching to account for observed covariates and show that peer influence effects are reduced by 300-700$\%$ afterward. Zamal et al. [12] find that augmenting Twitter user features with features derived from user friends significantly improves estimates of three assortative features: gender, age, and political affiliation.

One way to demonstrate peer influence effect bias, is to use ordinary least squares regression [3]. If there are $N$ data points, let $u_{it}$, the response variable of the $i$th person at time $t$, be modeled as:
\begin{equation} \label{eq:1}
    u_{it} = \beta_{intercept} + \beta_{peer}peer_{i,t-1} + \beta_{X} \cdot X_i + \beta_{Z} \cdot Z_i + \epsilon_{i}
\end{equation}
\noindent{}where $peer_{i,t-1}$ is the peer influence experienced by $i$ at time $t-1$. $X_i$ is vector of observed characteristics/demographics of $i$; $Z_i$ is a vector of unobserved characteristics. $(\beta_{intercept},\beta_{peer},\beta_{X},\beta_{Z})$ is the vector of linear regression coefficients for the response variable as a function of peer influence, observed characteristics, $X$, and latent characteristics, $Z$. If the latent characteristics are not represented in (\ref{eq:1}), then the resulting peer influence coefficient, $\beta_{peer}'$ will be biased ($\beta_{peer}' \ne \beta_{peer}$) if the latent characteristers are correlated with peer influence and if $\beta_Z \ne 0$. If latent characteristic proxies, $\hat{Z}$, are used and yield peer influence coefficent $\beta_{peer}''$, bias may be reduced ($|\beta_{peer}'' - \beta_{peer}| < |\beta_{peer}' - \beta_{peer}|$) if the proxies are a sufficient representation for $Z$. The objective of our study is to determine the effectiveness of latent homophily proxies in reducing peer influence effect bias.

\subsection{Latent variable proxies} \label{latent}

\textbf{Latent Coordinates}: We adopt the social latent space model developed by Hoff [5]. Given a social network defined by an adjacency matrix $A$ where $A_{ij} = 1$ if users $i$ and $j$ are friends/connected and 0 otherwise, observed demographics, $X_i$, unobserved characteristics, $\xi_i$, of person $i$, and intercept term, $\gamma_0$, the social latent space models the probability of two people being friends as:
\begin{equation} \label{eq:2}
    P(A_{ij}=1) = \frac{exp(\gamma_0 + |X_i-X_j| - |\xi_i-\xi_j|)}{1+exp(\gamma_0 + |X_i-X_j| - |\xi_i-\xi_j|)}
\end{equation}
We use the R package \textbf{vblpcm} to generate estimates of the coefficients, $\hat{\gamma_0}$, and the unobserved characteristics, $\hat{\xi}$, of dimension 2 through variational Bayesian inference [7]. We refer to $\left\{\hat{\xi}_i\right\}$ as latent coordinates, our first latent homophily proxy. This choice is justified since under (\ref{eq:2}), friends will tend to be closer in latent space than non-friends.

\textbf{Community membership}: Community membership is another candidate proxy for latent homophily [8]. People in the same community may have unobsered similarities that account for their relationships. Given a network, we infer community membership through Newman's fast community detection algorithm [6]. The algorithm works by initializing each node in a network as a community and agglomeratively combining communities that result in the biggest increase in modularity, which measures the strength of a network's community partition.

\section{Models}
\label{gen_inst}

 We formulate models under the assumption that data comes in samples with a binary response variable and predictor variables that can be continuous or discrete.

\subsection{Hierarchical logistic regression}

We adopt a hierarchical logistic regression algorithm [1] for modeling because it allows for joint inference of logistic regression parameters on the population and sample level and characterization of sample heterogeneity. Using the formulation in (\ref{eq:1}) but with a binary response variable, denote the logistic regression coefficients for a subset, $s$, by $\beta_s$. The hierarchical logistic regression algorithm assumes the following generative hierarchy for $\beta_s$: Given a data divided into $N$ subsets, the hierarchical logistic regression algorithm assumes that for each subset, $s$, the logistic regression coefficients, $\beta_s$, of dimension $n+1$ that describe the relationship between the response and $n$ predictor variables are generated from a common multivariate Normal distribution with mean $\delta$ and covariance $V_{\beta}$ [1]:
\begin{equation} \label{eq:3}
\beta_s \sim N(\delta,V_{\beta})
\end{equation}
\noindent{}The mean parameter, $\delta$, is assumed to be drawn from a Normal prior distribution with mean and covariance hyper parameters $\bar{\delta}$ and $A_{\delta}^{-1} \cdot I$, where $A_{\delta}$ is a scalar, respectively. The covariance parameter, $V_{\beta}$, is assumed to drawn from a Wishart distribution with $\nu$ degrees of freedom and covariance $V = \nu \cdot I$ where $I$ is the identity matrix of dimenion $n$: 
\begin{gather}
\delta \sim N\left(\bar{\delta},A_{\delta}^{-1} \cdot I \right)\\
V_{\beta}^{-1} \sim W^{-1}(\nu,V)
\end{gather}
The mean $\delta$ is a tuple of logistic regression coefficients characterizing the relationship between the response variable and predictor variables over the $N$ data subsets. The covariance matrix $V_{\beta}$ characterizes the heterogeneity of the logistic regression coefficients of the subsets. Hierarchical logistic regression returns samples of $\left\{\beta_s\right\}_{s=1}^N$, $\delta$, and $V_{\beta}$ estimated through Markov Chain Monte Carlo. We implement the algorithm [1] using the R package \textbf{bayesm}. We use default hyperparameter values $\bar{\delta}=0$, $\nu = n+3$, $A_{\delta} = 0.01$, take 4$\cdot 10^5$ samples with a thinning factor of 5 and a burnin of 2$\cdot 10^5$. Convergence is checked by visually examining trace plots. We take the posterior means of the samples of $\left\{\beta_s\right\}_{s=1}^N$, $\delta$, and $V_{\beta}$ as our estimates.

\subsection{Agglomerative clustering with regularized logistic regression}

Hierarchical logistic regression performs joint inference on data subsets assuming the subset logistic regression coefficients are drawn from a Normal distribution (\ref{eq:3}). We believe this to be a strong assumption since the logistic regression coefficients may display non Normal properties such as multimodality or skewness. To address this assumption, we develop a novel inference algorithm that agglomeratively concatenates compatible subsets and implements regularized logistic regression on the concatenated subsets to get regression coefficients. Compatibility is determined through an instability metric, instab for short, described in Algorithm 2. The algorithm is based on agglomerative clustering [11] and is described below:

\begin{algorithm}
\caption{Agglomerative clustering with regularized logistic regression}
\label{alg:alg2}
\begin{algorithmic}[1]
\REQUIRE A sample of $N$ data subsets, $\left\{s_i\right\}_{i=1}^N$, where each subset, $s_i$, contains $N_{s_i}$ instances of a binary response variable, $Y_{s_i}$, and corresponding features, $feat_{s_i}$
\ENSURE A sample of $N'$ clustered data subsets, $S' = \left\{s_i'\right\}_{i=1}^{N'}$, where each subset, $s_i'$, contains $N_{s_i'}$ instances of a binary response variable, $Y_{s_i'}$, and corresponding features, $feat_{s_i'}$. The regularized logistic regression coefficients of each $s_i'$
\STATE Combine subets until each contains at least 2 binary responses for each class, yielding $\left\{s_i'\right\}_{i=1}^{N'}$. Subsets that contain less than 2 binary responses of either class are merged according to minimum $T^2$ statistic, $T^2 = (\beta_i - \beta_j)^T(cov(\beta_i) + cov(\beta_j))^{-1}(\beta_i - \beta_j)$ where $\beta_i$ and $\beta_j$ are unregularized logistic regression coefficients for $s_i$ and $s_j$
\STATE $remInd \leftarrow 1,...,N'$
\WHILE{length($remInd$) > 1}
\STATE $(i,j) \leftarrow \text{argmin}\left\{\text{instab}\left(s_i' \cup s_j'\right) - \text{min}\left(\text{instab}(s_i'),\text{instab}(s_j')\right)\right\}$ 
\STATE $N' \leftarrow N' + 1$
\STATE $s_{N'} \leftarrow s_i' \cup s_j'$
\STATE children($s_{N'}$) $\leftarrow \left\{i,j\right\}$
\STATE $remInd \leftarrow remInd - \left\{i,j\right\} \cup N'$ 
\ENDWHILE
\STATE $splitQueue \leftarrow s_{remInd}'$
\STATE $S' \leftarrow \left\{\right\}$
\WHILE{$splitQueue$ is nonempty}
\STATE $s' \leftarrow$ dequeue($splitQueue$)
\STATE $(i,j) \leftarrow \text{children}(s')$
\IF{min(instab($s_i',s_j'$)) $<$ instab($s'$)}
\STATE $splitQueue.\text{enqueue}(s_i',s_j')$
\ELSE
\STATE $S' \leftarrow S' \cup s'$
\ENDIF
\ENDWHILE
\STATE $\beta_{s_i'} \leftarrow reglogistic(Y_{s_i'},feat_{s_i'})$, $i = 1,...,N'$
\ENSURE{$S',\left\{\beta_{s_i'} \right\}_{i=1}^{N'}$}
\end{algorithmic}
\end{algorithm}

We implement regularized logistic regression with a ridge penalty using the R package \textbf{glmnet}, represented by the command $reglogistic$ above. We use \textbf{glmnet} because it solves for coefficients, $\beta$,  according to 
\setlength{\arraycolsep}{0.0em}
\begin{eqnarray}
argmin_{\beta} \lambda\frac{1}{2}\beta^T\beta - [\sum_{i=1}^{N} (install_i)log(\hat{p}_i) + \nonumber\\  
(1-install_i) log(1-\hat{p}_i)] \nonumber
\end{eqnarray}
\setlength{\arraycolsep}{2pt}
for a wide range of $\lambda$ values. $\hat{p_i} = exp(\beta \cdot feature_i)/(1+exp(\beta \cdot feature_i))$, where $feature_i$ is the feature vector for the $i$th data point. We choose the $\lambda$ halfway between the minimum and maximum value to get moderate regularization. Since this requires at least 2 instances of both binary classes, we combine subsets until this criteria is met in line 1. In lines 3 to 9, we agglomeratively combines subsets until one remains. Subsets are combined to minimize instability, which we define according to Algorithm 2. We define instability of a data set as how sensitive the class 1 probabilities are to data variation. In lines 1-2, we first compute the class 1 probabilities using coefficients from all data points. We then create 100 ($N_p$) 90/10 ($f = 0.1$) partitions of the data in line 3. For each partition, the class 1 probabilities of the 10$\%$ held out data set using coefficients fit to the $90\%$ held in data set are computed in line 7. The mean absolute difference between those probabilities and the corresponding probabilities using coefficients from all data is computed in line 9. Returning to Algorithm 1, we then go the reverse direction and split the data into the component subsets to minimize instability in lines 12-19. This yields a partition of the subsets, $S'$, initialized in line 11. The outputs are $S'$ and the regularized logsitic regression coefficients, $\beta_{s_i'}$, for each cluster, $s_i', i = 1,...,N'$, of $S'$.  

\begin{algorithm}
\caption{Data instability}
\label{alg:alg3}
\begin{algorithmic}[1]
\REQUIRE A data set, $s$, with $N_{s}$ instances of a binary response variable (class 0 and 1), $Y_{s}$, and corresponding predictor variables/features, $feat_{s}$. The number of random partitions, $N_p$, to split $s$ into held in and held out subsets and the fraction of $s$, $f$, that is held out at each partition.
\ENSURE A measure of how instable $s$ is to data variation.
\STATE Compute regularized logistic regression coefficients for $s$: $\beta \leftarrow reglogistic(Y_{s}, feat_{s})$
\STATE Compute class 1 probabilities for each index, $x$, $x \in \left\{1,...,N_s\right\}$, $prob_x \leftarrow \frac{exp(\beta \cdot feat_{s,x})}{1+exp(\beta \cdot feat_{s,x})}$
\STATE Compute $N_p$ partitions of $s$ where each partition, $p$, consists of a set $I_{p} \subset \left\{1,...,N_s\right\}$ of $(1 - f)N_s$ indices representing held in data points and a corresponding set of $O_{p} = \left\{1,...,N_s\right\} - I_p$ held out indices. 
\STATE $instability \leftarrow array(N_s)$
\FOR{$p$ from 1 to $N_p$}
\STATE $\beta_{in} \leftarrow reglogistic(Y_s[I_p],feat_s[I_p])$
\STATE $prob_{out 1} \leftarrow \left\{\frac{exp(\beta_{in} \cdot feat_{s,x})}{1+exp(\beta_{in} \cdot feat_{s,x})} | x \in O_p \right\}$
\STATE $prob_{out 2} \leftarrow \left\{prob_x | x \in O_p \right\}$
\STATE $instability[p] \leftarrow mean(|prob_{out 1} - prob_{out 2}|)$
\ENDFOR
\ENSURE{$mean(instability)$}
\end{algorithmic}
\end{algorithm}

Analogous to the logistic regression coefficients seen in hierarchical logistic regression, $\left\{\beta_{s_i'}\right\}_{i=1}^{N'}$ represent the heterogenous relationships between the response and predictor variables seen throughout the data but on for the aggregated subsets rather than for the individual subsets. 

\section{Data}
\label{headings}

We use a week of data (October 13-19, 2010) from a popular Japanese social network website consisting of a complete social graph of all its members in the form of an edge list: 600 million edges among its 22 million users and a list of demographics for each user, including age, gender, number of comments, and number of photos. We define a snowball as a sample of website users that consists of a game installer and website users within 1 or 2 degrees of friendship from the game installer. Each user in a snowball is represented by a feature vector containing his/her unique, anonymous ID, proportion of his/her friends that downloaded or played the game the installer installed (nominal game) on the first six days, known as peer influence, demographic information, and a response variable of whether the user installed the nominal game on the seventh day. There are 15390 snowballs provided. We use the user feature vectors to predict game installation on the seventh day.

In order to focus on peer game usage in general and its effect on game installation we aggregate game installation information and peer influence across all games. We take the maximum of the user's set of peer influences as our representation of aggregated peer influence. Aggregating game installation and peer influence across games yields 7765 snowballs that just contain friends of the seed user and 1658 snowballs that contain users within 2 degrees of friendship from the seed user. We only use snowballs containing users within 2 degrees of friendship, which we now refer to as second degree snowballs, due to the richer data provided by the friends of friends of the seed user. 

In the second degree snowballs, peer influence is sparse: only 10$\%$ of website users have non-zero peer influence. To enrich the data, we select the second degree snowballs for which at least 15$\%$ of members have non-zero peer influence, yielding 318 snowballs. 

\subsection{Sampled Japanese website data} 
\label{real}
We select a random sample of 120 snowballs from the 318 target population snowballs (Tables I and II) to perform inference analysis through hierarchical logistic regression and the agglomerative model.
\begin {table}[H]
\centering
\caption{Summary of sampled snowballs with at least 15$\%$ users with non-zero peer influence}
\begin{tabular}{ll}
\toprule
    Filtered Data & Frequency \\ 
    \midrule
    Snowballs & 120 \\ 
    Website Users & 35278  \\ 
    Seventh Day Game Installations & 234  \\
\bottomrule
\end{tabular}
\end{table}

\begin {table}[H]
\centering
\caption{Summary statistics for sampled second degree snowballs with at least 15$\%$ users with non-zero peer influence}
\begin{tabular}{lllll}
\toprule
    Snowball Mean Feature & Mean & SD & Min & Max \\ 
    \midrule
    Installation ($\%$) & 1.2 & 1.4 & 0.10 & 11 \\
    Peer Activity ($\%$) & 0.55 & 0.42 & 0.16 & 2.4 \\ 
    Friends (log) & 4.3  & 0.54 & 2.3 & 6.1  \\
    Female & 0.57  & 0.15 & 0.13 & 0.96 \\
    Age & 24 & 3.8 & 17 & 35 \\
    Photos (log) & 2.4 & 0.73 & 0.50 & 4.1 \\
    Comments (log) & 2.3 & 1.0 & 0.10 & 4.0 \\
    Size  & 294 & 259 & 21 & 981 \\
\bottomrule
\end{tabular}
\end{table}
We augment the data with features reflecting peer characteristics and snowball structure (Table III).

\begin {table}[H]
\caption{Mean features (extended) of 120 sampled second degree snowballs with at least 15$\%$ users with non-zero peer influence}
\begin{center}
    \begin{tabular}{lllll}
    \toprule
    Extended Feature & Mean & SD & Min & Max \\ 
    \midrule
    Mean Gender of Friends (Female=1) & 0.58  & 0.095 & 0.28 & 0.81 \\
    Mean Age of Friends & 24 & 3.4 & 18 & 34 \\
    Mean Photos (log) of Friends & 2.6 & 0.56 & 1.3 & 3.8 \\
    Mean Comments (log) of Friends & 2.6 & 0.88 & 0.46 & 3.8 \\
    Distance from Friends & 12 & 17 & 1.4 & 87 \\
    Snowball Size & 294 & 259 & 21 & 981 \\
    Snowball Density & 0.060 & 0.049 & 0.0066 & 0.23 \\
    Number ofl Connected Components & 1.31 & 0.66 & 1 & 4 \\
    Latent Homophily & 0.31 & 2.7  & -5.9 &  5.6\\
    \bottomrule
    \end{tabular}
\end{center}
\end {table}
The distance of the user from his/her friends is defined as $\sum_{feature} \frac{user_{feature}-mean(friends_{feature})}{sd(friends_{feature})}$, where the features range over gender, age, number of photos (log), and number of comments (log). The number of connected components of a snowball excludes seed user and the seed user's edges since otherwise each snowball will be connected. We verified that the sampled snowballs cover a sufficient range of feature values of the target population. In order to analyze different subpopulations, we split the 120 snowballs into 12 samples of 10. For each of the 12 samples, we create ground truth homophily data in the form of a two dimensional latent coordinate for each snowball member via the variational Bayes model in \ref{latent}. We use the adjacency matrix containing all the website users in the sample and use gender, age, photos (log), and comments (log) as the input features. 

We define $\left\{\beta_{\text{true},s}\right\}_{s=1}^C$ as the logistic regression coefficients returned when either model is applied to a sample of 10 snowballs where the snowball members have observed (including extended) features and ground truth latent homophily, which we call ground truth snowballs. $C$ represents the number of snowballs in the case of the hierarchical logistic regression and the number of snowball clusters in the case of the agglomerative model. We define $\left\{\beta_{\text{naive},s}\right\}_{s=1}^C$ as the logistic regression coefficients returned when either model is applied to a sample of ground truth snowballs that have latent homophily omitted from the feature vectors. 

To test the effectiveness of latent coordinates as a latent homophily proxy, for each snowball we compute a two dimentional latent coordinate for each website member using the adjacency matrix for the snowball members and the same four features used to compute the ground truth latent homophily. We hypothesize latent coordinates computed individually for each snowball are a suitable proxy for ground truth latent homophily because they represent the localized network information. We thus refer to these as localized latent coordinates and the ground truth latent homophily as global latent coordinates. We define $\left\{\beta_{\text{latent},s}\right\}_{s=1}^C$ as the logistic regression coefficients returned when localized latent coordinates are used in place of global latent coordinates and $\left\{\beta_{\text{comm},s}\right\}_{s=1}^C$ as the logistic regression coefficients returned when community membership is used in place of global latent coordiantes. The coefficients will be used in computing peer influence effect bias.

\subsection{Simulated Japanese website data}
To test the effectiveness of the latent homophily proxies under different homophily effects, for each model, we create 9 copies, each with different latent homphily coefficients (\ref{eq:1}), of each of the 12 samples described in Table 1. Due to processing contraints, we could only simulate a limited number of conditions. We simulate snowballs by fitting the installation response data of each snowball, $s$, to a tuple of logistic regression coefficients, $\beta_{\text{generate},s}$. To keep the effects of the other predictor variables similar to those seen in the actual data, we set $\beta_{\text{generate},s} = \beta_{\text{true},s}$ for all predictor variable except peer influence and latent homophily. We set the peer influence coefficient for each snowball to 2 for simplicity. For a given sample, we then assign the latent homophily coefficients (remember there are two since the latent coordinates are two dimensional) in $\beta_{\text{generate},s}$ for each snowball in the sample to an ordered pair in (2,2), (2,-2), (-2,-2), (0,0), (0,1), (2,-1), (2,0), (2,1), (-2,0). We fit the installation data of a snowball to $\beta_{\text{generate},s}$ by setting the response $install_i = 1$ with probability $\hat{p_i} = exp(\beta_{generate} \cdot feature_i)/(1+exp(\beta_{generate} \cdot feature_i))$ where $feature_i$ is the feature vector for the $i$th user in $s$. Applying hierarchical logistic regression to each simulated sample yields ground truth coefficients $\left\{\beta_{\text{sim true},s}\right\}_{s=1}^C$. We then obtain coefficients $\left\{\beta_{\text{sim latent},s}\right\}_{s=1}^C$ and $\left\{\beta_{\text{sim comm},s}\right\}_{s=1}^C$ in the same way as in \ref{real}. Note $C$ = 10 since there are 10 snowballs per sample.

For the simulated snowballs to be used by the agglomerative model, $\left\{\beta_{generate,s}\right\}$ are inferred individually for each snowball using regularized logistic regression with a ridge penalty. We use the Python \textbf{sklearn} implementation of regularized logistic regression. It solves for the generating coefficients according to 
\setlength{\arraycolsep}{0.0em}
\begin{eqnarray}
argmin_{\beta} \frac{1}{2}\beta^T\beta - [\sum_{i=1}^{N} (install_i)log(\hat{p}_i) + \nonumber\\  
(1-install_i)log(1-\hat{p}_i)] \nonumber
\end{eqnarray}
\setlength{\arraycolsep}{2pt}
Unlike hierarchical logistic regression, the agglomerative model does not assume that each snowball in a sample has logistic regression coefficients generated from a Normal distribution. Thus snowball coefficients are fit independently. We assign the latent homophily coefficients of each snowball to the same ordered pair used for the hierarchical logistic regression simulated snowballs and fit the installation data as before. $\left\{\beta_{\text{sim true},s}\right\}_{s=1}^{C_{\text{true}}}$ and $\left\{\beta_{\text{sim comm},s}\right\}_{s=1}^{C_{\text{comm}}}$ are obtained with agglomerative clustering with regularized logistic regression. Note the number of snowball clusters obtained may change as the feature vectors of the website users change and that localized latent coordinates cannot be used a proxy since the cluster members change throughout the algorithm via merges and splits.

\section{Model performance}
\label{others}

\subsection{Cross validation}

For the hierarchical logistic regression algorithm (hierarchical LR), cross validation was implemented by using 50 randomly selected 90/10 samples. To test the robustness of the coefficients for each snowball in the sample, the absolute difference of the median probability of downloading a game for the training (90$\%$) and test (10$\%$) data as a percentage of the training median was computed. The median over the snowballs would represent the sample's average absolute difference between training and test data.

For the agglomerative model, we took the same 50 90/10 samples used for the hierarchical logistic regression model and implemented the same procedure described above using the agglomerative model. This time, however, the train and test data of the snowball clusters were compared. 

We implement cross validation on four samples of 10 ground truth snowballs (Table IV).
\begin {table}[H]
\title{Cross valdiation of hierarchical clustering and agglomerative model}
\caption{Median absolute difference between training and test data probability of game download}
\begin{center}
    \begin{tabular}{lllll}
    \toprule
    Algorithm & Sample 1 & Sample 2 & Sample 3 & Sample 4  \\ 
    \midrule
    Hierarchical LR & 100 & 100 & 100 & 114\\ 
    Agglomerative Model & 15 & 14 & 5.6 & 7.8 \\
    \bottomrule
    \end{tabular}
\end{center}
\end {table}
We use the unpooled t-test to compute the significance of the observed differences in medians between the algorithms for each sample. Agglomerative clustering proved to be significantly more robust than hierarchical logistic regression in all 4 samples at a critical level of 0.01. 

\subsection{Peer influence effect bias}
We study three different types of peer influence effect biases. The first is the bias of the peer influence effect when ground truth latent homophily is excluded from the data. This is calculated for each snowball, $s$. First, the peer influence effect for each member, $i$, of $s$ with non zero peer influence must be calculated by the following equation.
\begin{align} \label{eq:23}
    \frac{(p_{s,i}|\beta_{naive,s}) - (p_{s,i}|\beta_{naive,s},peer_i=0)}{|p_{s,i}|\beta_{naive,s}|} \cdot 100\%
\end{align}
$p_{s,i}|\beta_{naive,s} = e^{(\beta_{naive,s} \cdot features_{naive,i})}/(1+e^{(\beta_{naive,s} \cdot features_{naive,i})})$. $features_{naive,i}$ represents the full set of features with the exception of ground truth homophlily. (\ref{eq:23}) is the percent of game downloads attributable to peer influence for user $i$ when latent homophily is omitted. The median of this quantity over all members with non zero peer influence in $s$ is denoted by $peerInf_{s,naive}$. We then calculate the average peer influence effect in the true model using the following equation analogously to (\ref{eq:23}) but with the ground truth latent homophily. We denote this quantity $peerInf_{s,true}$. The peer influence effect bias of snowball $s$ is then calculated to be: 
\begin{align} \label{eq:25}
    bias_{s,naive} = \frac{peerInf_{s,naive}-peerInf_{s,true}}{|peerInf_{s,true}|} \cdot 100\%
\end{align}
(\ref{eq:25}) represents the percent error in peer influence effect when latent homophily is not used. The second bias we calculate is the bias of the peer influence effect when the ground truth latent homophily coordinates are replaced with localized latent coordinates. This is denoted by $bias_{s,latent}$ which is calculated analogously to $bias_{s,naive}$. The third bias we calculate is the bias of the peer influence effect when community membership is used as a proxy for latent homophily, which is denoted by $bias_{s,comm}$. 

To determine if there is significant improvement in bias if latent homophily proxies are used, we use the sign test since it is robust to outliers and there are outliers in the distribution of biases. We apply the sign test to each set of 12 samples. If 10 or more samples have a majority of snowballs (six or more out of ten) that display peer influence bias reduction when latent homophily proxies are used, then the bias reduction is significant at a critical level of 0.05 ($P(Binom(12,0.5) \ge 10) = 0.019$). Note for the agglomerative model, each snowball adopts the logistic regression coefficients of its cluster. Since there are nine scenarios tested for the simulation data, we apply the Bonferroni correction to get a new critical level of 0.05/9 = 5.6$\cdot 10^{-3}$. Thus in order for a homophily proxy's improvement to be significant, a simulated sample must have at least 11/12 ($P(Binom(12,0.5) \ge 11) = 3.2\cdot 10^{-3}$) snowballs display peer influence effect bias reduction. 

Under both models, we find no significant improvement or deterioration in peer influence effect bias in the simulated and real samples with either localized latent coordinate or community membership.
\begin{table}[H]
\caption{Number of snowballs with peer influence bias reduction and median unsigned bias}
\begin{center}
\begin{tabular}{lllll}
\toprule
    \textbf{Algorithm} & \textbf{Simulated} & \textbf{Real Data} \\ 
    \multicolumn{3}{r}{Improvement Ratio $\text{     }$}                    \\
    \cmidrule{2-3}
    \midrule
    Hierarchical LR with latent coord. & 550/1080 & 66/120\\
    Hierarchical LR with comm. mem. & 447/1080 & 54/120\\
    Agglomerative Model & $\textbf{666/1080}$ & 57/120\\ 
\bottomrule
\end{tabular}
\end{center}
\end{table}
\begin{table}[H]
\caption{Number of snowballs with peer influence bias reduction and median unsigned bias}
\begin{center}
\begin{tabular}{lll}
\toprule
    \textbf{Algorithm} & \textbf{Simulated} & \textbf{Real Data} \\ 
    \multicolumn{3}{r}{$|bias_{proxy}|$ ($|bias_{naive}|$)}                    \\
    \cmidrule{2-3}
    \midrule
    Hierarchical LR with latent coord. & 117 (116) & 173 (128)\\
    Hierarchical LR with comm. mem. & 144 (116) & 140 (128)\\
    Agglomerative Model & $\textbf{61 (93)}$ & 37 (16)\\ 
\bottomrule
\end{tabular}
\end{center}
\end{table}
Since we only have 12 samples, we did not have enough power to obtain significance with the sign test. However, since using community membership through the agglomerative model results in 666/1080 = 62$\%$ of simulated snowballs displaying peer influence effect bias reduction with a lower absolute bias on average (Table V), there seems to be potential with this model. We next turn to the peer influence effect bias seen in the real data. In Figure 1, we plot the distribution of biases for all 120 snowballs under both models when global latent coordinates are not used. 

\begin{figure}[H]
 \caption{Distribution of peer influence bias}
 \centering
   \includegraphics[width=1\linewidth]{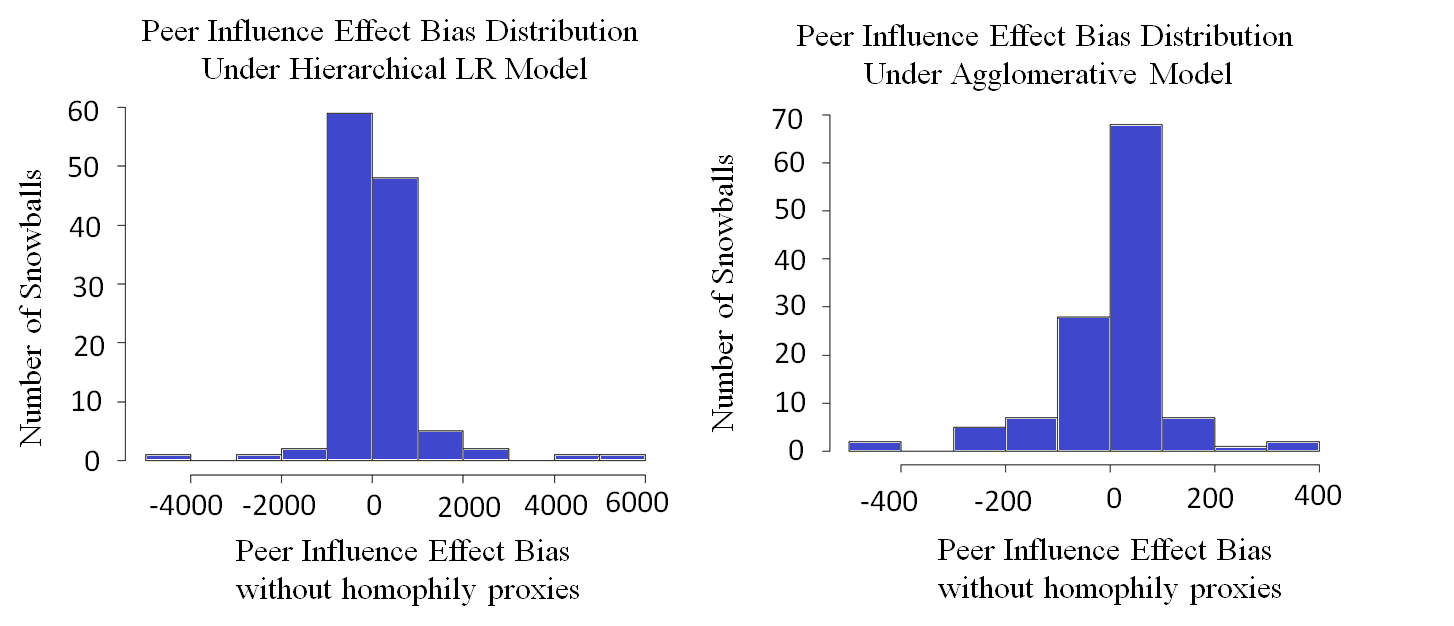}
\end{figure}

Under hierarchical logistic regression, the median absolute bias is 128$\%$ and under the agglomerative model the median absolute bias is 15.5$\%$. Latent homophily can result in a wide range of biases in both models, with both distributions having biases that exceed 100$\%$ in absolute value. When represented by global latent coordinates, latent homophily significantly impacts peer influence estimates. 
\section{Conclusion}
We derive a simulation framework for measuring peer influence effect bias improvement with latent homophily proxies. We model ground truth latent homophily with a latent space social network model. We implement the simulation framework on simulated data derived from a Japanese social network website and on real website data with two models for predicting game downloads. For the simulated data, we find no statistically significant bias reduction with latent homophily proxies under either model. Previous work [3] shows peer influence effect deflation through hierarchical logistic regression when localized latent coordinates are used as latent homophily proxies. However, we show peer influence effect estimates are not significantly improved with localized latent coordinates. We affirm this to be a critical finding since the peer influence effects in [3] may not necessarily be more accurate with latent coordinates. 

While significant bias reduction was not seen, this does not discount the usefulness of latent homophily proxies. Under the agglomerative model, 62$\%$ of simulated snowballs displayed bias reduction with community membership and the unsigned bias was less with community membership on average. Furthermore, we show the agglomerative model is more robust to data variation than hierarchical logistic regression when applied to the Japanese website data. This suggests that hierarchical logistic regression may not be a suitable model for the website data.

For the real data, we also find no statistically significant bias reduction under either model. However, we find that not representing latent homophily with global latent coordinates can result in extremely large changes in peer influence effect biases. Given that latent homophily can be reasonably modeled with global latent coordinates, these coordinates should be used when inferring peer influence effects in a network if processing contraints are met. If the network is too big, then less computationally demanding models of latent homophily should be used. However, we find localized latent coordinates and community membership increase bias in the real data on average when global latent coordinates are used as ground truth latent homophily. Thus for next steps we encourage investigation of alternative latent homophily proxies, conditions under which latent homophily effects are strong, alternative inference algorithms, and applications to other data sets.

\section*{References}

\small

\noindent{}[1] Allenby, G. GM., McColloch, R., Rossi, P.E. (2005). "Bayesian Statistics and Marketing", Wiley Series in Probability and Statistics, Appendix A.

\noindent{}[2] Aral, S., Muchnik, L., and Sundararajan, A. (2009)."Distinguishing influence-based contagion from homophily-driven diffusion in dynamic networks. Proceedings of the National Academy of Sciences", 106(51), 21544–21549.

\noindent{}[3] Davin, Joseph (2015). "Essays on the Social Consumer: Peer influence in the adoption and engagement of digital goods." PhD Thesis, Harvard Business School, 3-25.

\noindent{}[4] Hill, S., Provost, F., and Volinsky, C. (2006). Network-based marketing: Identifying likely adopters via consumer networks. Statistical Science, (pp. 256–276).

\noindent{}[5] Hoff PD, Raftery AE, Handcock MS (2002). "Latent Space Approaches to Social Network Analysis." Journal of the American Statistical Association, 97(460), 1090-1098.

\noindent{}[6] Newman, M. (2004). "Fast algorithm for detecting community structure in networks". Physical Review E. 69, 066133.  

\noindent{}[7] Salter-Townshend, M. and Murphy, T. (2013). "Variational Bayesian Inference for the Latent Position Cluster Model". The Tenth Annual Conference on Neural Information Processing Systems.

\noindent{}[8] Shalizi, C. R. and Thomas, A. C. (2011). "Homophily and contagion are generically confounded in observational social network studies." Sociological methods and research, 40(2), 211-239.

\noindent{}[9] Tucker, C. (2008). Identifying formal and informal influence in technology adoption with network externalities. Management Science, 54(12), 2024-2038.

\noindent{}[10] Van den Bulte, C. and Lilien, G (2001). "Medical Innovation Revisited: Social Contagion versus Marketing Effort." American Journal of Sociology. Vol. 106. No. 5. 1409-35.

\noindent{}[11] Ward, Joe H. (1963). "Hierarchical Grouping to Optimize an Objective Function". Journal of the American Statistical Association 58 (301): 236–244

\noindent{}[12] Zamal, F., Liu W., Ruths, D (2012). "Homophily and Latent Attribute Inference: Inferring Latent Attributes of Twitter Users from Neighbors." Proceedings of the Sixth International Association for the Advancement of Artificial Intelligence Conference on Weblogs and Social Media.

\end{document}